\documentclass[a4paper,12pt]{article}
\usepackage[T2A]{fontenc}
\usepackage[cp1251]{inputenc}
\usepackage[english]{babel}
\usepackage{graphicx}
\usepackage{graphics}
\usepackage{amssymb}
\usepackage{amsmath}

\begin{document}

\begin{titlepage}
\begin{center}
~\\
\vspace{1cm}~\\

{\bf \Large Simulation of the Polarized Fermion Decays} \\
\vspace{2cm} A.A.~Ashimova~\footnote{E-mail:~ashimova\_aa$@$mail.ru} \\
{\it Moscow State University,\\
 Moscow 119992, Russia} \\[2mm]
S.R.~Slabospitsky~\footnote{E-mail:~Sergey.Slabospitsky$@$ihep.ru}\\
{\it  State Research Center\\
Institute for High Energy Physics,\\
Protvino, Moscow Region 142281, Russia}

\end{center}

\vspace{1cm}
\begin{abstract} \noindent
In this paper the modification of the method conventionally used
for the modeling of the massive fermions production and decays is proposed.
The step by step algorithm is presented. Under the
strict conditions the proposed method of modeling allow distinctly
raise the efficiency of the computations.
\end{abstract}
\end{titlepage}

\newpage

Simulation of massive fermion production (let us denote it $F$)
with subsequent  decay frequently occurs in High Energy Physics.
Indeed,  the angular distributions of the fermion decay products
are very sensitive to its polarization. 

\begin{figure}[h]
 \centering \includegraphics{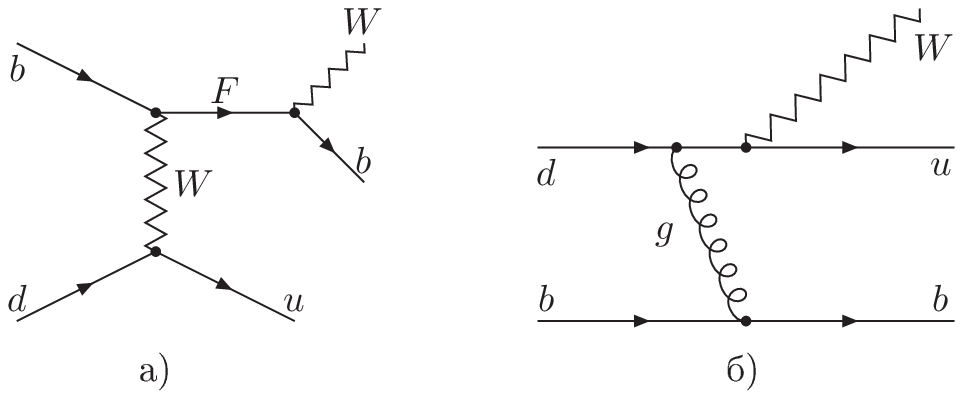}\\
 \caption{\small\it The example of the diagrams describing the process 
with the same final state.
                    Diagram a) corresponds to the massive fermion $F$ 
resonance contribution. Here $F$ decays into $bW$.
                    Diagram b) has  the same final
                    state, but not contains fermion~$F$.}
 \label{fig1}
\end{figure}

The most obvious way is the calculation of the
squared amplitude of the whole process including the fermion decays.
 However such way may be accompanied by some
difficulties. First, the calculation of the amplitude of the whole process
(including the decay chain) could be rather cumbersome. More over
 the corresponding final state can
appear without the massive fermion contribution. In other words
the final state of the process could be the result of the
additional diagrams not containing $F$ at all (see Fig.~\ref{fig1}).

However, one can assume that the invariant mass of the $F$ decay
products is close to fermion mass $m_F$. In this case the contribution of
non-resonance diagrams (see Fig.\ref{fig1}b) will be notably
suppressed. 

Fairly often the problem could be simplified with the 
narrow resonance approximation, 
i.e. the intermediate fermion is considered to
be <<on the shell>> ($p^2_F=m^2_F$, $p_F$ is 4-momentum of the
fermion $F$). Such approach allows to simplify the calculations
and split the simulation into two steps: 
\begin{enumerate}
    \item the production of the massive polarized fermion;
    \item the subsequent decay of the polarized fermion 
\end{enumerate}
Thereby the squared amplitude of the process can be written in the 
<<factorized>> form:
\begin{equation}\label{factor}
    |A|^2=|A^{Prod}(s)|^2\cdot|A^{Dec}(s)|^2\otimes \Phi(s).
\end{equation}
Here $A^{Prod}(s)$ and $A^{Dec}(s)$  are the amplitudes describing the 
production and decay of the polarized fermion, $s$ denotes the spin of
the fermion ($(sp_F)~=~0$) and $\Phi(s)$ is the factor of <<spin transfer>>
 from production to decay.

The problem posed in such way used to be solved within the method
of spiral amplitudes \cite{Richardson}. In this method the
amplitudes of the fermion $F$ production and decay calculated
depending on the different helicities ($\lambda_i$). Than the
total amplitude of the process can be performed as the following
sum:
\begin{equation}\label{MSA}
    |A|^2=\sum_{\lambda_i,\lambda_j}A^{Prod}(\lambda_i)A^{Dec}(\lambda_i)
 \cdot (A^{Prod}(\lambda_j))^+(A^{Dec}(\lambda_j))^+,
\end{equation}

   The equivalent approach was proposed by Jadach and Was \cite{Jadach}.
They showed that expression (\ref{MSA}) can be rewritten as
follows:
\begin{equation}\label{hv}
    |A|^2=|A^{Prod}_0|^2\cdot|A^{Dec}_0|^2(1+\vec{H}\vec{V}).
\end{equation}
Here $A^{Prod}_0$ and $A^{Dec}_0$ are the amplitudes of {\bf
unpolarized} fermion production and decay. $\vec{H}$ and $\vec{V}$
are so-called polarization vectors. They are determined in the fermion
rest frame and contain information about fermion spin.
\begin{eqnarray}
  |A^{Prod}(s)|^2 &=& |A^{Prod}_0|^2\left(1+(Hs)\right)  \label{hs}\\
  |A^{Dec}(s)|^2 &=& |A^{Dec}_0|^2\left(1+(Vs)\right)  \label{vs}
\end{eqnarray}

One example is given below. 
Accordingly to (\ref{vs}) the matrix element squared for the
$t$~quark decay to $bW^+$ with the following decay of $W^+$~boson
to $\ell\nu$ can be written as follows:
\begin{equation}\label{tbW}
 |M|^2 = 2\frac{(p_b p_{\nu}) (p_{t} p_{\ell}) }
 {(p^2_{w} - M_W^2)^2 + \Gamma_W^2 M_W^2}
   \times     \left[1 - \frac{m_t (p_{\ell} s)}{ (p_{t} p_{\ell})} \right],
\end{equation}
where $p_t, p_b, p_l, p_nu$ are the momenta of the $t$ and $b$ quarks and
final leptons, $\Gamma_W$ and $M_W$ are the total decay width and the mass
of the $W$-boson.
Hence one can derive the polarization vector for this decay
\begin{equation}\label{VtbW}
     V^{\mu} = -\frac{m_t p^{\mu}_{\ell}}{(p_t p_{\ell})}
\end{equation}

\begin{figure}[h]
  \centering \includegraphics[width=6.cm,height=2.5cm,clip]{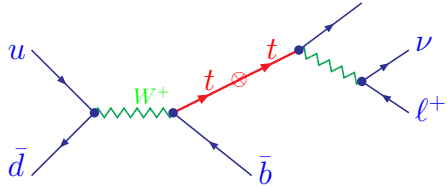}\\
  \caption{\small\it Feynman diagram for the process $u \bar{d} \to b \ell 
\nu \bar b $}
  \label{prim}
\end{figure}

As an example for expression~(\ref{hv}) the process $u \bar d \; \to \;
t \bar b$, $t \to b \ell \nu$ is considered (see Fig.\ref{prim}).

Using (\ref{tbW}) the explicit expression for the matrix element squared 
of  this process can be written in the form (compare to~(\ref{hv})) 
as follows:
\begin{eqnarray}
 |M|^2 &\propto& \left( g^4 \frac{(p_u p_{\bar b}) (p_{t} p_{\bar d})}
 {(p^2_{w1} - M_W^2)^2 + \Gamma_W^2 M_W^2} \right) \times  \nonumber \\
&&\left( g^4 \frac{(p_b p_{\nu}) (p_{t} p_{\ell})}
 {(p^2_{w2} - M_W^2)^2 + \Gamma_W^2 M_W^2} \right) \times
 \left( 1 + \vec{n}_{\ell}~\vec{n}_{\bar d} \right),
\end{eqnarray}
where $\vec{n}_{\ell}$ and $\vec{n}_{\bar d}$ are directions of
$\ell^+$ and $\bar d$-quark momenta in $t$-quark rest frame.

Fairly often the expression (\ref{hv}) is preferable in the
numerical calculations since it reduces the quantity of logical
operation. Such conventional method of simulation uses discarding
technique (the reject-and accept method)~\cite{Nummeth}
 and supposes realization of the following
algorithm (see details in \cite{Jadach}):
\begin{enumerate}
\item [1)] to calculate the filal momenta of the particles from
  the  fermion $F$ production process;
\vspace{-4mm}
 \item [2)] according to a given kinematics to evaluate the polarization 
 vector  $\vec{H}$ in $F$ rest frame;
\vspace{-4mm}
 \item [3)] to perform simulation of the $F$-fermion  decay and to fix the 
 decay products kinematics;
\vspace{-4mm}
 \item [4)] to evaluate the decay polarization vector $\vec{V}$ 
  (in $F$ rest frame);
\vspace{-4mm}
 \item [5)] to calculate the additional weight $W = (1+\vec{H}\vec{V})$
    and using a discarding technique to reject or accept an event;
\vspace{-4mm}
    \item [6)] if the event is discarded than return to step 3).
\end{enumerate}
However using this algorithm one can expect the increase of
computation time in case of complicated expressions for the
massive fermion matrix elements.

In this Note the modification of this  algorithm is
proposed. Very often such modification allows to raise
an efficiency of the numerical calculations. The basic condition
of the modified algorithm application is  $|\vec{V}|=const$, i.e.
the vector $\vec{V}$ absolute value must be independent of the
fermion decay simulation results.

One should notice that the value of
the additional weight $W = (1+\vec{H}\vec{V})$ (see step 5 of the
algorithm) depends on the direction of $\vec{V}$ only.
 Besides the different kinematics of the
decay can lead to the same polarization vector $\vec{V}$. This
ideas are the basic for our modified algorithm. 

Thus we offer to chose correctly the polarization vector $\vec{V}$
(actually the direction of $\vec{V}$) before the fermion decay
simulation. Correctly means in order to accept an event. Hence the
algorithms steps starting from step 3) modifies:

\begin{enumerate}
 \item [3as)] chose the $\vec{V}$ direction using an discarding technique 
 with the weight  $W=1+\vec{H}\vec{V}$;
\vspace{-4mm}
    \item [4as)] to simulate the massive fermion decay ;
\vspace{-4mm}
    \item [5as)] to evaluate the polarization vector $\vec{v}$ corresponding 
  to the $F$ decay kinematics
    (mote, that in general $\vec{v}\neq \vec{V}$ since the vectors 
 may have different directions);
\vspace{-4mm}
    \item [6as)] to rotate the  reference system so that vector $\vec{v}$ 
 coincides with $\vec{V}$ in $F$ rest frame .
\end{enumerate}
\vspace{-4mm}

Let us point out that for each accepted event the simulation of
fermion decay made just {\bf once}. This fact is for sure an
advantage of the proposed algorithm.
Note, that the method described above can be laso used in  case
of two fermions production with a subsequent decays. In this case 
the additional weight takes the form:
\begin{equation}\label{W2}
    W_2 = 1 + \vec{h}_1\vec{v}_1 + \vec{h}_2\vec{v}_2 +
    h_{ik}v^i_1v^k_2 .
\end{equation}
where (like in~(\ref{hv}))  two vectors $h_{1,2}$ and tensor $h_{ik}$ are
determined by production kinematics.
The simulation of such two fermions decays and the turning of the
reference systems should be made independently.

\vspace{4mm}
The authors appreciate greatly Z.~Was for the plentiful discussions
and thanks to S.K.~Abdullin, V.V.~Kabachenko, A.G.~Miagkov,
A.~de~Roeck for the debates and censorious remarks.

This work was supported in part by Russian Foundation for Basic
Research under Grant \# 08-02-91002-C


\end{document}